\documentclass{article}

\usepackage{arxiv}

\usepackage[utf8]{inputenc} 
\usepackage[T1]{fontenc}    
\usepackage{hyperref}       
\usepackage{url}            
\usepackage{booktabs}       
\usepackage{amsfonts}       
\usepackage{nicefrac}       
\usepackage{microtype}      
\usepackage{lipsum}		
\usepackage{graphicx}
\usepackage{natbib}
\usepackage{doi}
\usepackage{cleveref}
\usepackage{multirow}

\title{Knee arthritis severity measurement using deep learning: a publicly available algorithm with a multi-institutional validation showing radiologist-level performance}

\author{Hanxue Gu
\And
Keyu Li
\And
Roy J. Colglazier
\And
Jichen Yang
\And
Michael Lebhar
\And
Jonathan O'Donnell
\And
William A. Jiranek
\And
Richard C. Mather
\And
Rob J. French
\And
Nicholas Said
\And
Jikai Zhang
\And
Christine Park
\And
Maciej A. Mazurowski \\
\thanks{Michael Lebhar is with the Duke University of Medicine.}
\thanks{Roy J. Colglazier, Nicholas Said, Rob J. French are with the Department of Radiology, Radiology, Musculoskeletal Imaging, Radiology.}
\thanks{Jonathan O’Donnell is with the Practice Transformation Unit in the Department of Orthopaedic Surgery at Duke University.}
\thanks{William A. Jiranek is with the Orthopaedic Surgery, Orthopaedics, Orthopaedics at Duke University.}
\thanks{Richard C. Mather is with the Orthopaedic Surgery, Orthopaedics, Orthopaedics; Population Health Sciences, Population Health Sciences, Basic Science Departments; Duke Clinical Research Institute, Duke Clinical Research Institute, Institutes and Centers; Duke-Margolis Center for Health Policy, Duke - Margolis Center For Health Policy, Initiatives.}
\thanks{Maciej A. Mazurowski is with Radiology, Radiology, Clinical Science Departments; Biostatistics and Bioinformatics, Biostatistics \& Bioinformatics, Basic Science Departments; the Department of Electrical and Computer Engineering, Electrical and Computer Engineering, Pratt School of Engineering; Computer Science, Computer Science, Trinity College of Arts \& Sciences; Duke Cancer Institute, Duke Cancer Institute, Institutes and Centers in Duke University, Durham 27708, United States.}
}

\renewcommand{\shorttitle}{\textit{arXiv} Template}

\hypersetup{
pdftitle={A template for the arxiv style},
pdfsubject={q-bio.NC, q-bio.QM},
pdfauthor={David S.~Hippocampus, Elias D.~Striatum},
pdfkeywords={First keyword, Second keyword, More},
}

\begin{document}
\maketitle

\begin{abstract}
	The assessment of knee osteoarthritis (KOA) severity on knee X-rays is a central criteria for the use of total knee arthroplasty. However, this assessment suffers from imprecise standards and a remarkably high inter-reader variability. An algorithmic, automated assessment of KOA severity could improve overall outcomes of knee replacement procedures by increasing the appropriateness of its use. We propose a novel deep learning-based five-step algorithm to automatically grade KOA from posterior-anterior (PA) views of radiographs: (1) image preprocessing (2) localization of knees joints in the image using the YOLO v3-Tiny model, (3) initial assessment of the severity of osteoarthritis using a convolutional neural network-based classifier, (4) segmentation of the joints and calculation of the joint space narrowing (JSN), and  (5), a combination of the JSN and the initial assessment to determine a final Kellgren-Lawrence (KL) score. Furthermore, by displaying the segmentation masks used to make the assessment, our algorithm demonstrates a higher degree of transparency compared to typical ``black box'' deep learning classifiers. We perform a comprehensive evaluation using two public datasets and one dataset from our institution, and show that our algorithm reaches state-of-the art performance. Moreover, we also collected ratings from multiple radiologists at our institution and showed that our algorithm performs at the radiologist level.

The software has been made publicly available at
\textit{https://github.com/MaciejMazurowski/osteoarthritis-classification.}
\end{abstract}


\section{Introduction}
\label{sec:introduction}
Osteoarthritis (OA) \cite{1} is a common form of joint disorder, with knee OA being a leading cause of disability among older adults. Knee osteoarthritis (KOA) typically manifests itself as the loss of articular cartilage from the medial and/or lateral femoral condyles as well as the patellofemoral articular surfaces. This can be idiopathic, caused by mechanical overload or post-traumatic changes. 
KOA can also manifest with the development of subchondral sclerosis, osteophytes, and periarticular cyst formation \cite{Guccione1994,Vos2012,Helmick2008}.

Knee joint replacement surgery is a safe and effective treatment to address functional deficits, pain and deformities \cite{5}.  However, significant concerns of this procedure exist, and as many as 20\% of patients remain dissatisfied after knee replacement for multiple reasons, such as mental health issues and degenerative physical function \cite{Nakano2020}.
According to The Agency for Healthcare Research and Quality, more than 790,000 knee replacements are performed each year in the US \cite{6}. Based on a new report of medical claims data from 2010-2017 by the Blue Cross Blue Shield Association (BCBSA), the average price for an inpatient’s knee replacement is \$30,249, the number of knee replacements is up by 17\%, and the average cost increased by 6\% during this period \cite{7}. Furthermore, due to an aging population with a longer life expectancy \cite{Losina2012}, there is a projected 143\% increase in  total knee arthroplasty (TKA) by 2050 \cite{Inacio2017}. 
This combination of a significant dissatisfaction rate, surgical risks, increased utilization and high expenses have fueled an increased focus on determining when knee replacement surgery is appropriate \cite{Porter2010}. 

However, by using published appropriateness criteria, one study \cite{Riddle2015} found that about one-third of TKA in the US was ``inappropriate'', and only about half were ``clearly appropriate''. 
The radiographic evaluation of KOA severity plays an important role in several TKA appropriateness criteria that have been proposed \cite{12, Riddle2014}, with knee replacement being indicative of more advanced diseases.  
Different criteria have been proposed for this purpose, with Kellgren-Lawrence (KL) grade assessment emerging as an effective tool that incorporates a radiologist’s measurement of joint space narrowing (JSN), osteophyte description and bony changes. The KL grade scales from 0 to 4, and is correlated with increasing severity of OA \cite{KELLGREN1957}. 
KL assessment was originally performed using anteroposterior (AP) knee radiographs, but recently demonstrated a greater sensitivity and specificity on posteroanterior (PA) views \cite{Meyer2017}. 

Unfortunately, KL assessment is associated with significant inter-reader variability, i.e., non-negotiable differences of grades among readers, which diminishes its wide applicability \cite{Sun1997,Riddle2013,Klara2016}. 
To improve the process of KOA assessment while also solving the aforementioned issues, the research of computer-aided assessment methods has increased in recent years. 
In contrast to approaches based mainly on manual visual feature extraction, deep learning (DL) has recently shown revolutionary success in imaging subspecialties, such as object detection \cite{Ren2015,Cai2016} and face recognition \cite{Lin1997,Lawrence1997}, due to the automated learning of features directly from data.
In addition, a growing number of attempts to apply DL models to medical images have achieved decent results, indicating the possibility of using these powerful tools in clinical practice in the near future. Indeed, multiple deep learning-based methods have recently been proposed for OA assessment \cite{Tiulpin2018,Antony2017,Swiecicki2021,kondal2020automatic,1500491,biology10111107,Selvaraju2016,saleem_x-ray_2020}. However, most of them rely solely on models that automatically provide a graded output from the radiographs, but this decision procedure is difficult to interpret, and often considered to be a ``black box''. For example, the method of \cite{Selvaraju2016}  adopts the Grad-CAM method for knee radiographs, and shows prediction heatmaps from a multitask deep learning model, but it cannot provide healthcare stakeholders with the precise information needed to help diagnostic procedures since it only shows which features help the model's decision, such as emphasizing joint regions, but not helping to analyze those features. In clinical practice, transparency of decision procedures is crucial for practitioner use and validation \cite{Haibe-Kains2020}.
On the other hand, there are methods \cite{1500491,biology10111107, saleem_x-ray_2020} that forgo classification directly in the original image for the sake of interpretability of OA evaluation, and instead select features commonly used by radiologists as the basis for OA evaluation, such as the most commonly used JSN. 
These methods are proven to be effective to some extent, but using selected features alone often leaves out other referable conditions in the image, such as bony changes, which making the judgments less informative and the performance relatively worse.

In our study, we propose a deep learning-based algorithm for improved accuracy and consistency of KL score assessment, while also improving transparency by providing a segmentation mask generated by the algorithm which is used for determining the KL score as well as a joint space narrowing grading. By showing the segmentation results at the knee joint, we can clearly display the morphological characteristics of the knee joints; at the same time, visualizing the distance of the joint, we can more intuitively infer the degree of knee osteoarthritis. We evaluate our method with several datasets, including the publicly available MOST datasets \cite{MOST}, the OAI dataset \cite{OAI}, and a Duke institutional dataset that gathered from Duke Health
System. We further provide assessments of several radiologists on these datasets, in order to validate our model.
In summary, our contributions include:
\begin{enumerate}
\item We propose and implement a novel algorithm that combines classification and segmentation in parallel to arrive at a more accurate measurement of KL grade. Our algorithm also results in a higher degree of transparency, as it provides a measurement of joint space narrowing and displays the segmentation mask that was used for assessment.
\item We provide a thorough multi-institutional evaluation of our method with three datasets, including two publicly available datasets and one collected at the Duke Health System. We demonstrate state-of-the art performance on all datasets, and are the first to our knowledge to provide this comprehensive of an evaluation.
\item We collect KL grade ratings from multiple radiologists on two of the datasets to evaluate the agreement between our method and radiologists. We compare to multiple radiologists to mitigate the problem of inter-reader variability, which is crucial to establish a clear baseline for our algorithm. We are the first in the literature to demonstrate this on these datasets for this task.
\item We share our software publicly as both Python code and executable software with a graphical user interface. We anticipate that this will significantly aid research on osteoarthritis assessment and surgery appropriateness criteria, and make our algorithm more widely accessible.
\end{enumerate}

\section{METHODS}
\label{sec: methods}
In this section, we first detail the preparation of the datasets and the corresponding annotations needed for training and validation, and then we introduce our KL grading classification algorithm consisting of five main steps in \Cref{sec: Algorithm}.
\begin{figure*}[]
\centerline{\includegraphics[width=\textwidth]{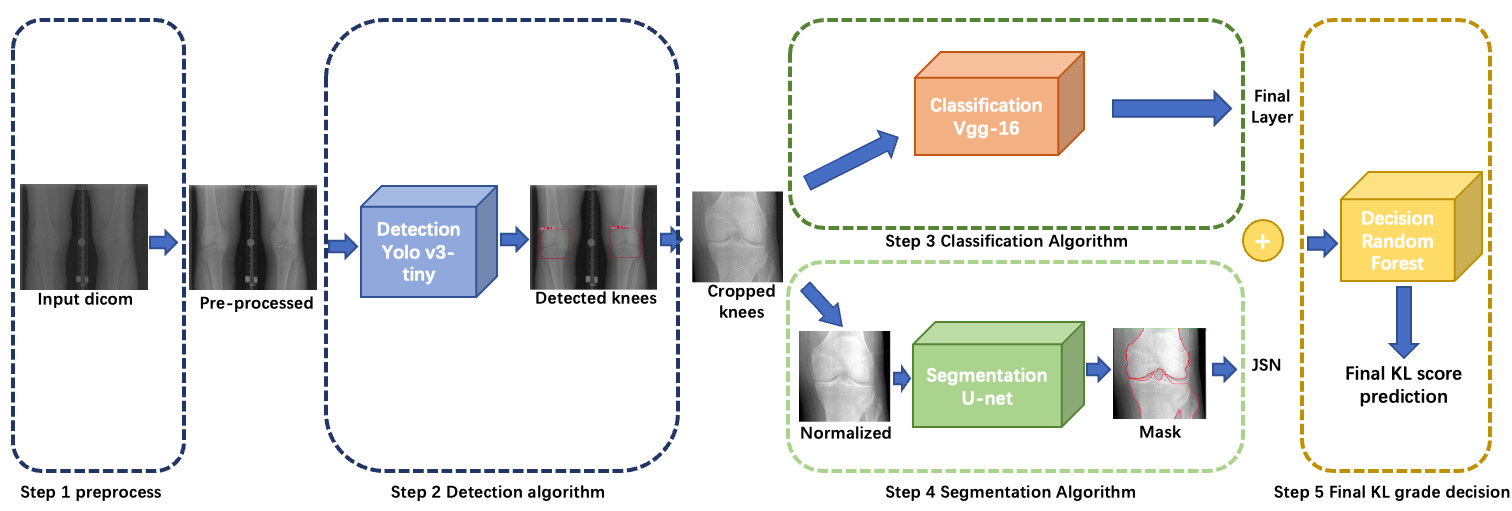}}
\caption{KOA severity prediction pipeline. The software has been shared publicly (https://github.com/MaciejMazurowski/osteoarthritis-classification).}
\label{fig1}
\end{figure*}

\subsection{Dataset}
\subsubsection{Dataset 1} 
We built our algorithm and conducted the first phase of testing based on the dataset from the Multicenter Osteoarthritis Study (MOST) \cite{MOST}, which collected longitudinal knee radiological data and clinical assessments. The original dataset consists of 10,052 exams from 3,036 patients, and it includes Kellgren-Lawrence grades, joint space narrowing (JSN) scores and osteophyte grade assessments. 
Radiologists assign Kellgren-Lawrence grades according to the following rules \cite{Mildenberger2002}:
\begin{enumerate}
\item Grade 0: no radiographic features of OA are present.
\item Grade 1: doubtful joint space narrowing (JSN) and possible osteophytic lipping.
\item Grade 2: definite osteophytes and possible JSN.
\item Grade 3: multiple osteophytes, definite JSN, sclerosis, possible bony deformity.
\item Grade 4: large osteophytes, marked JSN, severe sclerosis and definite bony deformity.
\end{enumerate}
To create a subset from the dataset for the analyses, we used the following inclusion criteria: 
\begin{enumerate}
\item If a knee at a visit was specifically marked as unfit in the dataset (for example, due to missing data), this was not considered in our analysis for that visit. 
\item We excluded images lacking either JSN or osteophyte grades. 
\item We excluded all visits with missing Posterior-Anterior (PA) or lateral (LAT) views. For visits with multiple PA views, we randomly select one view.
\end{enumerate}
Following these exclusions, our dataset for analysis contained 9,739 exams from 2,802 patients. This translated to 18,053 knees with 9,739 PA images. We randomly split these 2,802 patients into non-overlapping training, validation, and test subsets for basic classification, shown in \Cref{table1}, it follows the same criteria of \cite{Swiecicki2021}. The distribution of respective KL grades is presented in \Cref{table2}. In the next sections, we further describe how to annotate and prepare the detection, segmentation, and classification datasets needed for training and the test datasets for validation as well as additional datasets annotated by our institutional radiologists.

\begin{table}[]
\setlength{\tabcolsep}{2.5pt}
\caption{The prepared dataset 1.}
\label{table1}
\begin{tabular}{p{95pt}p{45pt}p{45pt}p{45pt}}
\hline
Classification dataset  & Training & Validation & Test \\ \hline
Number of patients     & 2040     & 259        & 503  \\
Number of exams        & 7062     & 914        & 1763 \\
Number of knees        & 13404    & 1740       & 3359 \\ \hline
\end{tabular}
\end{table}

\begin{table}[]
\setlength{\tabcolsep}{6pt}
\caption{The distribution of different KL grades for our dataset 1 (\Cref{table1}.}
\label{table2}
\begin{tabular}{p{52pt}p{25pt}p{25pt}p{25pt}p{25pt}p{25pt}}
\hline
\multirow{2}{*}{Set} & \multicolumn{5}{c}{KL grade}     \\ \cline{2-6} 
                     & 0    & 1    & 2    & 3    & 4    \\ \hline
Train                & 5600 & 1951 & 2228 & 2475 & 1150 \\ 
Valid                & 678  & 263  & 329  & 316  & 154  \\ 
Test                 & 1283 & 526  & 588  & 697  & 265  \\ \hline
\end{tabular}
\end{table}
\noindent \textit{1)	Detection annotations.} 300 PA view images (300 left and 300 right knees) were randomly selected from the training set of \textit{dataset 1}. These 300 images were then randomly divided into 210 images for training of the detection model, 45 images for the validation of the model, and 45 images for the testing of the detection model. Three experienced researchers under our institution annotated the center point of each knee joint as the key-point for detection.

\noindent \textit{2)	Segmentation annotations.} 600 PA view images were selected from the classification dataset to develop and evaluate our segmentation algorithm. We randomly selected 400 images from the training set, and 72 images from the validation set, of \textit{dataset 1}. Researchers from our institution manually annotated the segmentation masks based on software OsiriX \cite{osirix}, which included skeletal contours of the fibula, tibia, and femur.

\noindent \textit{3)	Kellgren-Lawrence grade annotations by radiologists from our institution.} In addition to the KL score annotations available in the MOST dataset, we randomly selected 110 PA view images from the test set of \textit{dataset 1} for additional assessment of the KL grade by 5 radiologists at our institution.
These images were graded by 4 board-certified Radiology Fellows/Clinical Instructors and 1 attending Radiologist with 5 years post-fellowship experience in the Musculoskeletal Division of the Duke Department of Radiology.

\subsubsection{Dataset 2}
To demonstrate the stability and generalization ability of our algorithm, we also utilized another publicly available knee dataset, The Osteoarthritis Initiative (OAI)  \cite{OAI}. OAI is a longitudinal study of 4796 participants examined with X-ray, MRI, and other metadata during nine follow-up examinations (0-96 months). Our \textit{Dataset 2} come from the OAI released screening package 0.C.2. After filtering out cases which lack JSN or KL grading, there are 2529 subjects with 5028 knees remaining. To be noticed, MOST dataset and OAI dataset covered subjects aged 50-79 and 45-79, respectively, which introduces different OA distributions across all subjects, and the different composition of data sources also introduced various image acquisition artefacts. 

\subsubsection{Dataset 3}
For the Duke dataset, 200 cases are selected from the electronic medical record (EMR) database at Duke in 2019 and the corresponding DICOM objects were retrieved by querying a large institutional picture archiving and communication system (PACS). The 200 images were graded on an assessment of KL grade by 3 radiologists at our institution. It is worth noting that these two datasets are only applied for the model performance testing stage and play no rule for algorithm construction.

\subsection{Algorithm}\label{sec: Algorithm}

Our method has five steps: 

\begin{enumerate}
    \item Step 1: Image preprocessing to resize the pixel spacing and normalize the image intensity.
    \item Step 2: Joint detection algorithm to localize the joint in the entire image and prepare a bounding box for future cropping.
    \item Step 3: Classification algorithm, which makes an initial guess which offers the probability for each grade. 
    \item Step 4: Segmentation algorithm and assessment of JSN level based on the segmentation masks. This proceeds in parallel with step 3.
    \item Step 5: Combining the information from the third and the fourth step, to make a final KL grade decision. 
    
\end{enumerate}

\Cref{fig1} shows the entire pipeline, and we also share the software.


\begin{figure*}[h]
\centerline{\includegraphics[width=0.78\textwidth]{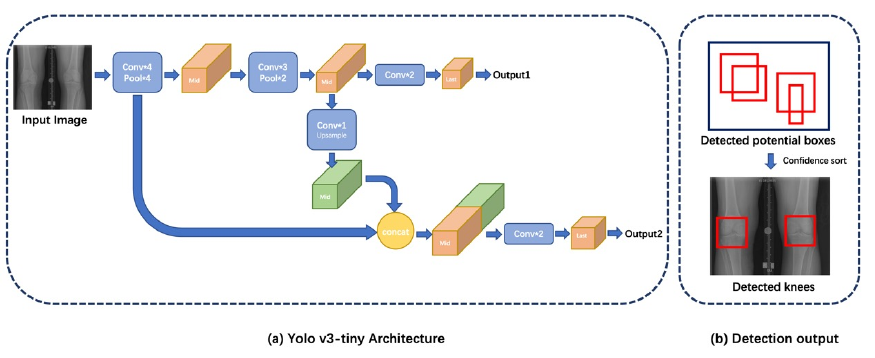}}
\caption{Joint detection model using Yolo v3-tiny. (a) Yolo v3-tiny network structure (b) the protocol of getting two detected knees.}
\label{fig2}
\end{figure*}

\subsubsection{Step 1: Image preprocessing} 
We use the``PixelSpacing'' tag from the DICOM images and the bicubic interpolation algorithm to resize the images into a uniform resolution, 0.2mm per pixel, which is a common value for radiographs. Then we convert all 16-bit grayscale images into 8-bit grayscale images. Finally, we normalize the pixel intensities by dividing all pixels by 255.

\subsubsection{Step 2: Joint Detection Algorithm} 
\label{sec:alg step2}
To provide a focused region of interest (ROI) and decrease downstream computations, we utilize a detection model to automatically localize the knee joints beforehand. Our detection model is built based on the Yolo V3-Tiny structure \cite{Adarsh2020}, and the detected bounding boxes are utilized to get the centers of joints. Yolo V3-Tiny is a simplified model from the well-known detection model Yolo V3, which implements two-scales of Yolo output layers instead of Residual layers. Yolo V3-Tiny achieves a speed three orders of magnitude higher than R-CNN, and two higher than Fast R-CNN, which could improve time efficiency for future usage of our approach. Our entire detection model is shown as \Cref{fig2}.

During training, we provided bounding boxes from previously annotated images, which were centered on the knee joint and were 500 pixels high and wide. The traditional loss function in Yolo V3 is a combination of position-prediction loss, size-prediction loss, confidence loss and classification loss. For our single-class object detection model, the last component only distinguishes the background out. To improve the model’s generalization ability, random rotations of up to 15 degrees were applied at each batch. 

During the inference process, the confidence level of proposed boxes was in great danger of changing dramatically as the image quality of this large dataset varies, which might incur an uncertain amount of detected knees and hinder our next process. Therefore, instead of controlling detected objects through confidence thresholds, we ranked the detected boxes for each image by confidence, and in which we select the top two as the final detected knees. By this method, we could consistently detect two knees even in low-quality images. 

\subsubsection{Step 3: Classification algorithm} We employed a convolutional neural network-based classifier, which accepts images of joints as inputs and automatically assign KL grades to them. The network backbone is based on the VGG-16 network \cite{Simonyan2014}, with added batch-normalization before max-pooling layers. 

The inputs to the classification network were the cropped images of the knee joint predicted by the detection model described in \Cref{sec:alg step2}. It is worth noting that in order to preserve the resolution and scale of the original image, we did not crop the image based on the detected box but cut a fixed patch of 672*672 pixels centered around the detected knee joint, and scaled it to 256*256 pixels to match the input of the VGG-16 network. Also, since there is a imbalance between classes \cref{table2}, we adopted a balanced-sampling to get rid of class bias during training.

\subsubsection{Step 4: Segmentation and joint space narrowing assessment} 
When the cartilage no longer keeps the bones a normal distance apart, joint space narrowing arises \cite{Braun2012}. JSN is a crucial marker of osteoarthritis and plays an instrumental role in the assessments of the KL grade. For this step, we assessed JSN through deep learning-based bone segmentation and an automatic measurement of the distance between bones.

\textit{Step 4a Segmentation algorithm.} With the same input as the classification algorithm mentioned before, after detecting the joint position based on the detection algorithm, we obtained a square with an edge length of 672 pixels extracted from the center of the detected knee joint.

\begin{figure*}[h]
\centerline{\includegraphics[width=0.8\textwidth]{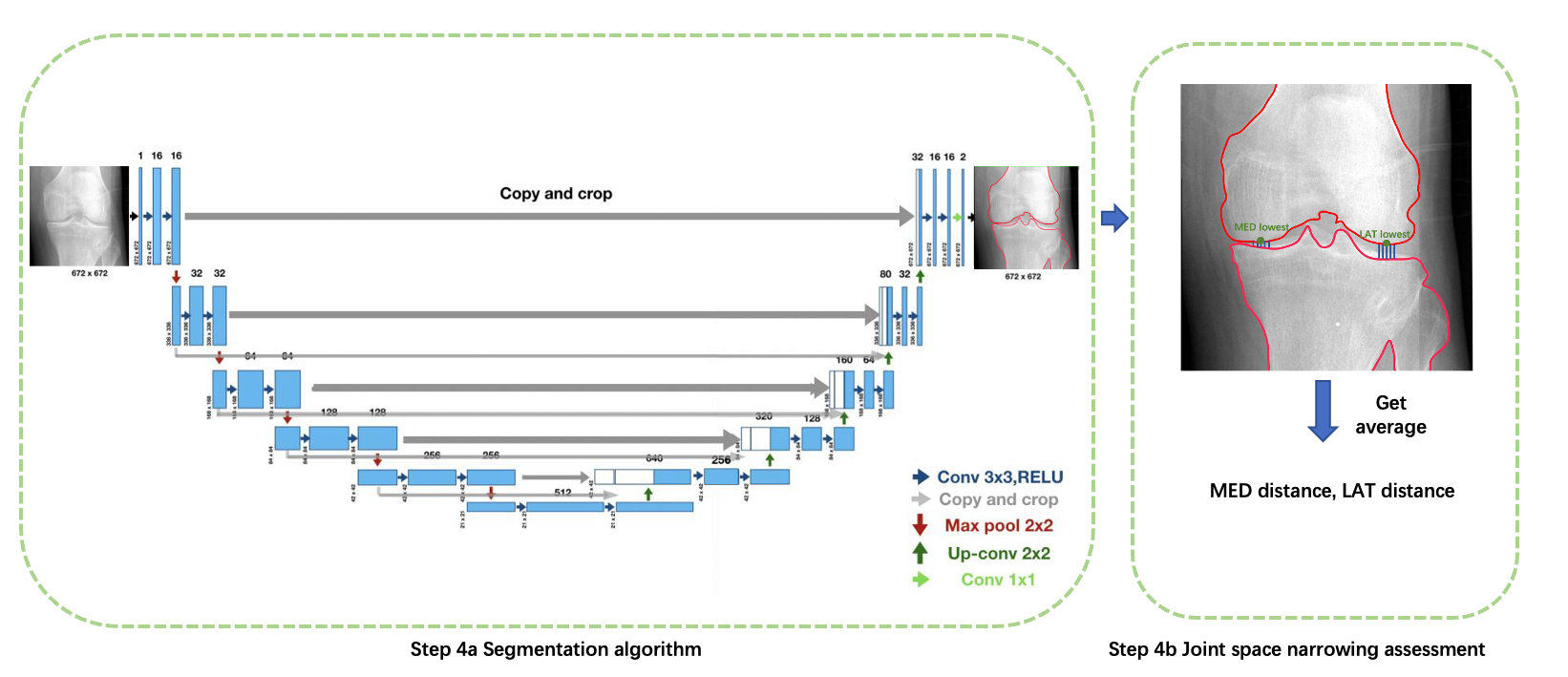}}
\caption{Segmentation algorithm and the joint space narrowing assessment. For step 4b, the green points are the lowest points on upper bone for LAT side and MED side; the blue lines are the generated series of vertical lines.}
\label{fig4}
\end{figure*}

In this step, we adopted the U-net segmentation architecture \cite{Ronneberger2015} with Coordinate Convolution layers \cite{Liu2018} as the model for the bone segmentation, which is a well-established method in medical image segmentation models in recent years. The network architecture is shown in \Cref{fig4} (step 4a).
We set the input size of the U-net to $672\times672\times3$ with 3 RGB channels after converting our joint images from gray to RGB. We also divided each knee annotation into two objects to be treated separately in the segmentation process: upper bone (femur) and the bone below (we regarded fibula and tibia as one object). 

\begin{figure}
\centerline{\includegraphics[width=0.8\columnwidth]{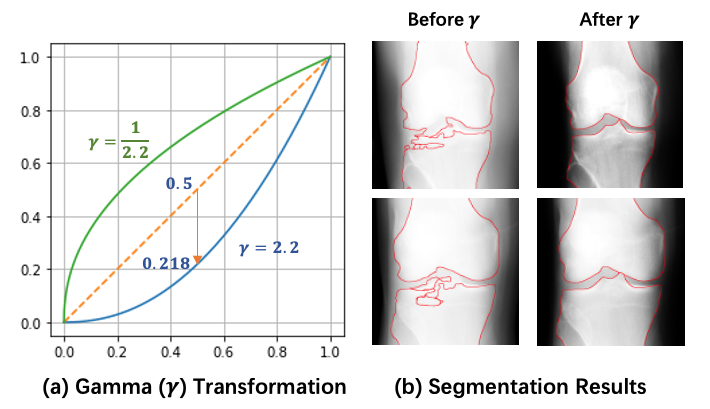}}
\caption{gamma correction, (a) The algorithm diagram; (b) the segmentation performance before and after adding gamma transformation.}
\label{fig5}
\end{figure}

Some images suffer from poor quality, thus, we applied a gamma-correction during evaluation, making our bones more distinguishable. As \Cref{fig5} (a) shown, when taking gamma $\gamma>1$, it provides a wider range for the lighter parts, making bones more detectable. Also, gamma is automatically adapted by the average intensity of the $50\times50$ box above the joint center (which must occur within the upper bone). Based on this gamma transformation, the dice coefficient on the evaluation set was improved. \Cref{fig5} (b) shows a significant improvement in segmentation accuracy, especially at the articular cartilage, which is the area of interest. After gamma correction, Laplacian sharpening, a well-established method for sharpening image edges \cite{Gupta2016}, was also applied, and a sharpening ratio of 30\% was used. These image enhancements are only applied during the inference stage, as we found based on experiments that adding the above enhancements reduces the model's generalization ability for low quality images, which affects the overall performance.

\textit{Step 4b Joint space narrowing assessment.} The segmentation algorithm above provided two masks: one for the upper part of the joint and one for the lower part. 
\begin{figure}[h]
\centerline{\includegraphics[width=0.8\columnwidth]{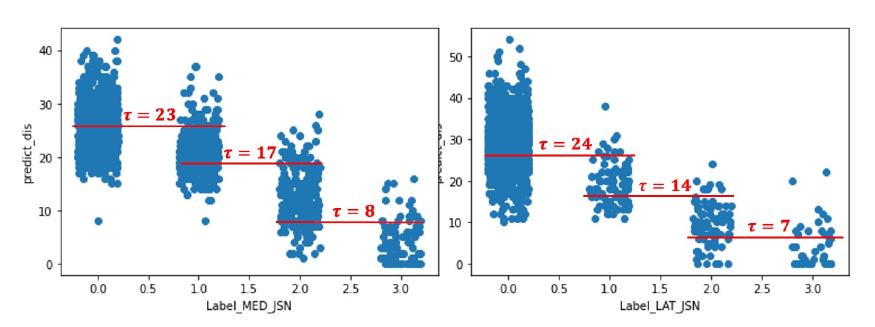}}
\caption{Diagram of joint space narrowing, (left) is the transfer mapping for medial knee, x-axis is the label for medial JSN, and y-axis is the calculated joint space distance. (right) is the transfer mapping for lateral knee. x-axis is the labels for JSN and y-axis is the predicted joint space distance.}
\label{fig6}
\end{figure}

The joint space narrowing is defined based on the MED (the inner part of joints), JSN and LAT (the external part of joints) side. We calculated the joint space distance (JSD) with the following method: (1) find the two lowest points for the upper bone, one is on the MED side and the other is on the LAT side, shown as green points in \Cref{fig4} (step 4b); (2) generate a series of vertical lines between the femur and tibia near the lowest points, shown as blue lines in \Cref{fig4} (step 4b); (3) get the average length $D_{avg}$  of these lines for MED and LAT side respectively.

After obtaining the JSD, k-means clustering \cite{Bock2007} is applied to complete the JSN degree assessment, which is based on the JSN ground-truth labels provided by the MOST dataset.
Specifically, by calculating the average MED joint space distance and the average LAT joint space distance of all the pairs on the evaluation set, we will get 1740 pairs of distance.
If by dividing them into 4 classes and minimizing the intra-class variance, we can obtain decision boundaries $tau$, 8, 17, 23 for the inner side and 7, 14, 24 for the outer side, with the transformed labels \Cref{fig8}  (left) and (right).
It is important to note that these assessment labels are only used for performance evaluation of JSN grading and for practitioners' reference.  We will directly pass the $2 \times 1$ JSDs (MED and LAT sides) to the next step.

\begin{figure*}
\centerline{\includegraphics[width=0.8\textwidth]{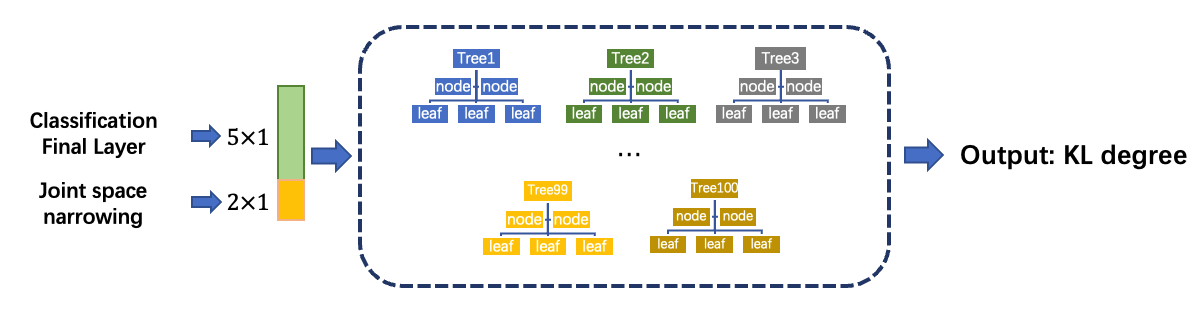}}
\caption{Procedure of Random Forest for the final KL-grade decision making.}
\label{fig7}
\end{figure*}

\subsubsection{Step 5: Decision making algorithm} 
At this point of the algorithm, we obtain a classification output in the form of a probability vector representing the possibilities of each class in step 3, and in step 4 we obtain the JSD from the segmentation model.
In this step, we merge these two vectors and adopt them to the final evaluation of the KL grades. Specifically, connecting these two vectors gives us a $7 \times 1$ vector, where we then use them as input to the final random forest classifier,  \cite{Breiman2001} to make a refined decision.
The random forest algorithm (RFA) (\Cref{fig7}) is an ensemble learning method for classification, which is a combination of decision trees and the output is a vote of all the individual trees.
During training, bootstrap sampling is applied to reduce correlations between trees so that each tree acts like an independent expert (analogous to a committee of radiologists), and our forest contains a set of 100 decision trees, each with a maximum depth of 8.
To improve time efficiency and avoid overfitting, the minimum number of sample leaves per branch is limited to two and the bagging method is used in the fitting process.
This large set of relatively uncorrelated models (trees) receive inputs simultaneously and operate as a committee to vote for the final decision, which we consider as a strategy to reduce inter-reader instability.

\section{MODEL EVALUATION AND RESULTS}
\label{sec: eva}
\subsection{Model Evaluation}
\subsubsection{Evaluation of the knee detection algorithm}
To evaluate our detection algorithm, we applied the commonly applied Intersection over Union (IoU) metric on the detection test set of the MOST dataset (Dataset 1). 
The IoU is the intersection area of the annotated box and the predicted joint knees’ bounding box divided by the union area of the two boxes. 
Since our annotation is based on the joint center \Cref{sec:alg step2}, we also computed the Euclidean distance $C_d$ between the labeled center and our predicted center, and called it the predicted deviation distance, and computed the average and maximum deviation distances on the detection dataset to evaluate the stability of the detection algorithm.

\subsubsection{Evaluation of the bone segmentation algorithm} 
The dice similarity coefficient, also known as the Sorensen-Dice index or simply as the dice coefficient, is a statistical metric for measuring the similarity between two sets of data.
We employed this metric to evaluate segmentation performance, as it is the most commonly employed evaluation metric for segmentation algorithms. We conducted the evaluation using the validation set of the MOST dataset (dataset 1). 

\subsubsection{Evaluation of the joint space narrowing assessment algorithm} 
To evaluate the JSN grading performance, we included two metrics for evaluation: accuracy and the confusion matrix. 
The ground truth JSN labels were annotated from the MOST dataset (dataset 1).  
Confusion matrix, in which the number of correct and incorrect predictions are summarized by count values and subdivided by each class, provides a visualization of the performance of the classification algorithm.
\begin{figure}[]
\centerline{\includegraphics[width=0.8\columnwidth]{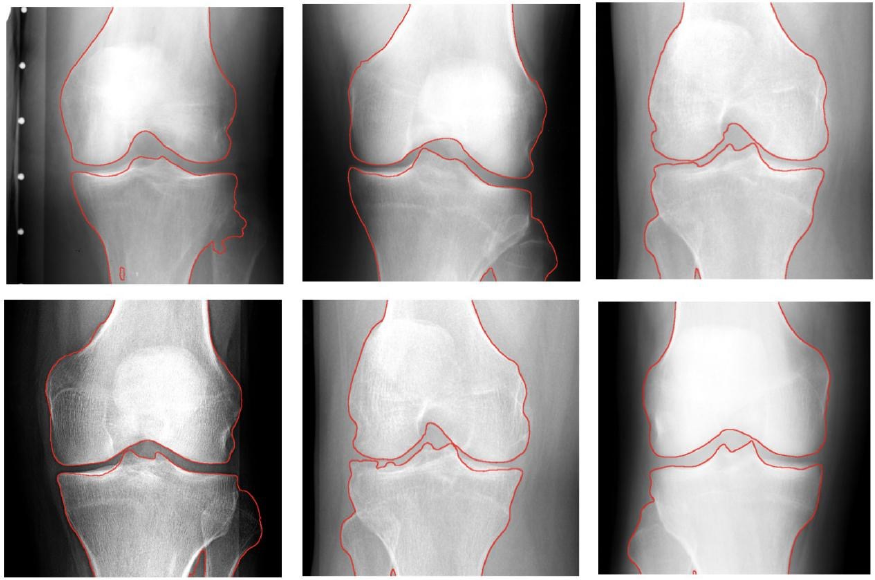}}
\caption{Segmentation results on the test set, the red line is the outline of the predicted mask.}
\label{fig8}
\end{figure}

\subsubsection{Evaluation of the complete algorithm for assessing KL grades using Dataset 1 and Dataset 2}
For the final classification model, in addition to the basic KL classification accuracy and confusion matrix, we also included the average recall and accuracy to provide more comprehensive evaluation details.
We compared the classification results with other state-of-the-art methods that also use OAI or MOST as the final evaluation dataset; classification accuracy and multi-class balance accuracy were chosen as the final metrics, which are shared by the different methods separately \cite{Antony2017,Tiulpin2018,Thomas2020,Swiecicki2021}. 
For a fair comparison, we displayed both the MOST test results and the OAI test results of our algorithm. 
As described in Section 2, the MOST test set was collected as a separate subset consisting of 1763 examinations, while the OAI test set was the entire collection of 2763 examinations.
From previous studies, the performance in detecting the presence of radioactive OA was also an important aspect of the KL grading method. A KL score of 2 was of exceptional significance as it was often used as a threshold for determining the occurrence of OA. 
Following the studies of \cite{Antony2017,Tiulpin2018,Thomas2020}, we combined KL scores of 0 and 1 as one class (negative) and KL scores of 2, 3 and 4 as another class (positive) to determine the presence of OA, \textit{i.e.,} $KL \geq 2$ represents the presence of OA. 
Precision, F1-score and classification accuracy were chosen as metrics for this binary OA instance analysis task.

\subsubsection{Evaluation of the complete algorithm compared with multiple radiologists on Dataset 1 and Dataset 3} 
To evaluate our models’ performance compared with radiologists, we introduced the coefficient $\kappa$ to evaluate the agreement between the two labeling methods, where $\kappa \leq 0$ indicates no agreement, and 1.0 means perfect agreement. 
The evaluation was based on 110 cases selected from the MOST dataset (Dataset 1) as described in section 2 A 3) and 200 selected cases from the Duke dataset (Dataset 3) as described in section 2 A 4). 
For the MOST dataset, $\kappa$ among the 5 radiologists, our ML predictions and the labels provided by MOST were all calculated separately, without stipulating which instance was the ground truth.
For the Duke dataset, since ground truth labels are lacking, we only compare the performance of our model with the 3 radiologists grading by $\kappa$.

\subsection{Results}
\subsubsection{Knee Joint Detection} The detection algorithm achieved an average IoU of 95.42\% for right knees and 95.12\% for left knees (\Cref{table3}). Among the testing images, all 90 joints were successfully detected, and the center of the detected joints ($C_d$) are all within a mismatch distance of 30 pixels, within less than 2\% of the image scale.

\begin{table}[]
\centering
\caption{Detection IoU on test set}
\label{table3}
\begin{tabular}{p{95pt}p{60pt}p{60pt}}\hline
         & Right Knee        & Left Knee      \\ \hline
IoU mean (\%) & 95.42 +/- 3.10  & 95.12+/-2.81 \\ 
$C_d$ mean (pixels) & 8.52 +/- 0.31 & 8.41 +/- 0.25 \\ 
$C_d$ max (pixels) & 26.01 +/- 2.32 & 23.79 +/- 3.43 \\ \hline
\end{tabular}
\end{table}

\subsubsection{Bone Segmentation} 
Our segmentation performed with an average Dice coefficient of 98.78\%. Some examples of segmentation results are provided in \Cref{fig8}; the presented images are pre-processed to a better quality (\Cref{sec: Algorithm}: step 4) and the red curve is the outline of our segmentation masks. Note that our segmentation algorithm can detect bone contour details, especially on the challenging joint region, which is crucial for accurate joint space distance calculation.

\subsubsection{Joint Space Narrowing} 
\begin{figure}[]
\centerline{\includegraphics[width = \columnwidth]{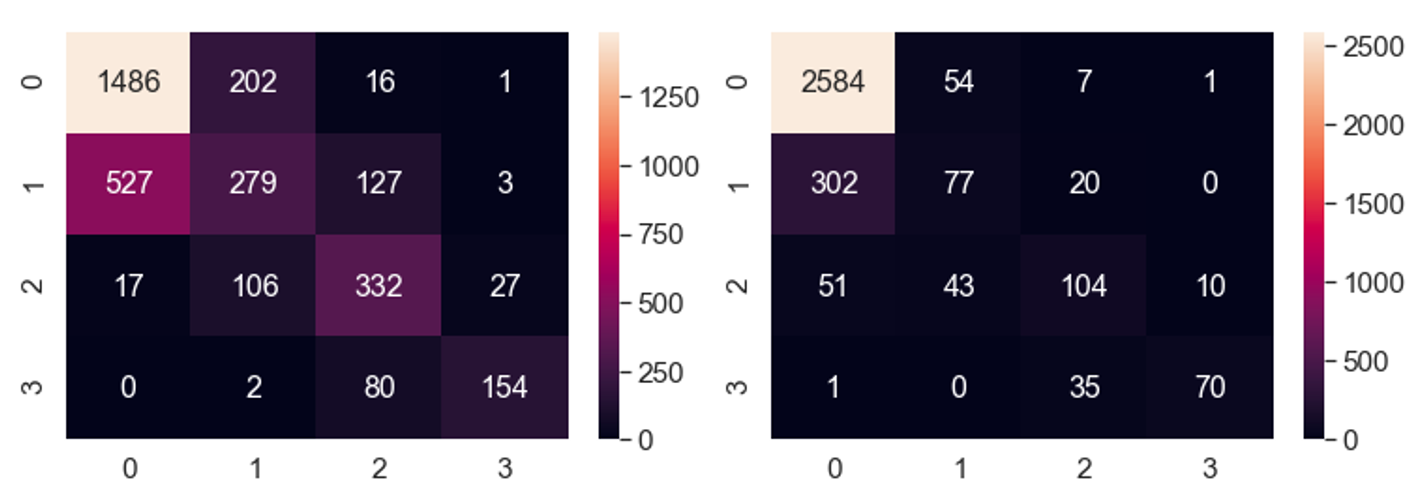}}
\caption{JSN classification confusion matrix, (left) JSN for medial knee; (right) JSN for lateral knee.}
\label{fig9}
\end{figure}

After transforming the measured joint space distances into JSN grades, our JSN classification algorithm yielded 67.01\% accuracy on medial knees and 84.40\% on lateral knees on the MOST test set, as shown in \Cref{table4}. Multi-class weighted F1-scores and precisions are 0.65, 0.65 for the medial side and 0.84, 0.81 for the lateral side, respectively. Confusion matrices are shown in \Cref{fig9}.

\subsubsection{Improvement of KL Score Prediction.} 
\begin{figure}[]
\centerline{\includegraphics[width=\columnwidth]{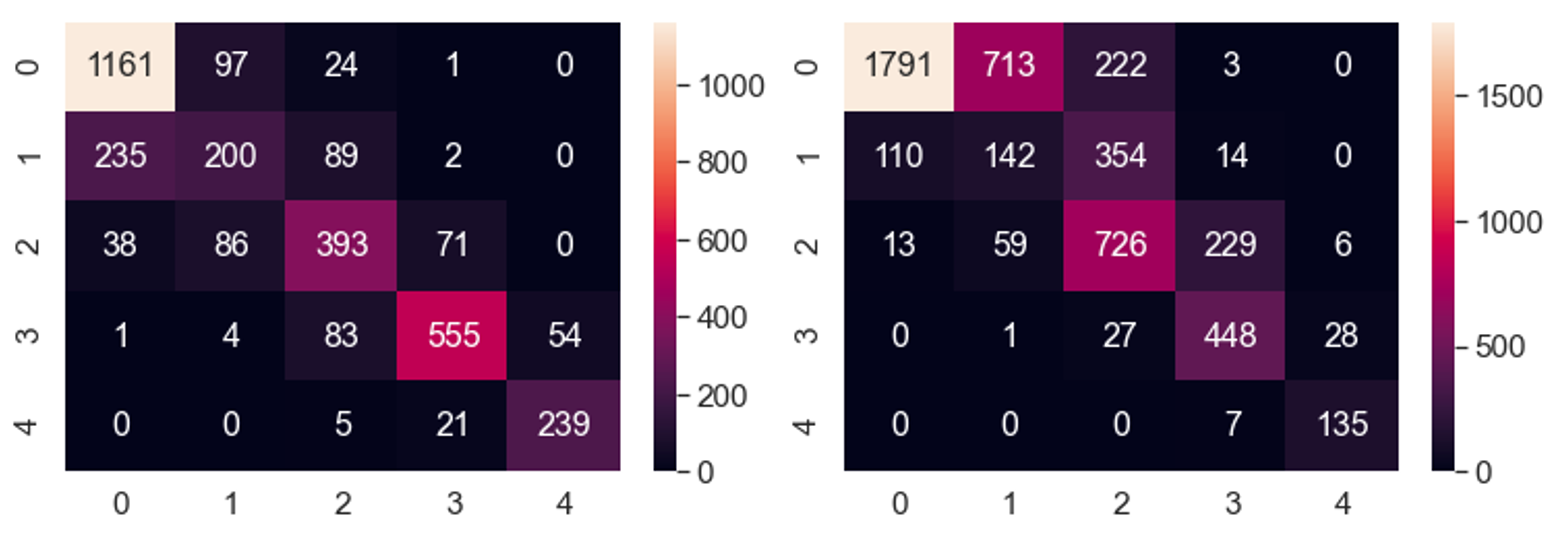}}
\caption{Confusion matrix for final KL classification results. (left) for MOST (dataset1) test set; (right) for OAI dataset (dataset 2).}
\label{fig10}
\end{figure}

The final model yielded \textbf{75.86\%} average accuracy and a \textbf{73.11\%} multi-class balanced accuracy on the entire MOST test set, and it obtained a mean class-wise precision of 0.73 and a mean class-wise recall of 0.73. The confusion matrix is shown in \Cref{fig10} (a), and it indicated that 93.52\% of the mis-graded knees were off by only one grade. When the deep learning-based classification model was used alone (without the JSN component), the average accuracy for the single classification model (no segmentation applied) was 71.93\%. The accuracy of the model that was based on JSN alone was 58.86\%. Besides the testing result on the MOST dataset (dataset 1), the average accuracy on OAI dataset (dataset 2) is 64.48\% and balanced accuracy is 68.56\%. The confusion matrix is shown as \Cref{fig10} (b).
The detailed comparison of classification performance with the state-of-art algorithms is shown in \Cref{table5}. 

For binary OA instance analysis, we achieved 0.92 for the precision, 0.94 for the F1-score and 0.93 for the average classification accuracy on the MOST dataset (dataset 1), and 0.89 for precision, 0.87 for F1-score and 0.87 for average classification accuracy.
\begin{table*}[]
\centering
\caption{Joint Space Narrowing Prediction on MOST (dataset 1) test set}
\label{table4}
\begin{tabular}{p{95pt}p{60pt}p{60pt}}
\hline
                   & Medial side & Lateral side \\ \hline
Predicted Accuracy & 67.01\%     & 84.40\%      \\ 
F1-score & 0.65 & 0.84 \\ 
Precision & 0.65 & 0.81 \\ \hline
\end{tabular}
\end{table*}

\begin{table*}[]
\caption{Comparison of classification results with other methods}
\label{table5}
\begin{tabular}{p{70pt}p{40pt}p{40pt}p{40pt}p{70pt}p{100pt}p{25pt}p{25pt}}
\hline
Study  & Train data & Val   data & Test data  & Detection  & Methodology & Acc. & Balanced Acc. \\ \hline
Antony,   et al., 2017 \cite{Antony2017}  & MOST+OAI    & MOST+OAI        & MOST+OAI & Manual                               & Classification   only                                                    & 61.90\% & - \\
Antony,   et al., 2017           & MOST+OAI    & MOST+OAI       & MOST+OAI & Faster R-CNN                         & Classification   and Regression                                          & 63.60\%  & -\\
Tiulpin,   et al., 2018 \cite{Tiulpin2018} & MOST          & OAI               & OAI        & Faster R-CNN with manual corrections & Separate   channels for lateral and medial compartments and model fusion & - & 66.71\%  \\
Tiulpin,   et al., 2018  \cite{Tiulpin2018}         & MOST          & OAI               & OAI        & Faster R-CNN with manual corrections & Fine-tuned   pretrained ResNet-34 (Transfer learning)                    & - & 67.49\%  \\
Thomas   et al., 2020  \cite{Thomas2020} & OAI           & OAI               & OAI        & Unknown                              & Fine-tuned pretrained ResNet-169 (Transfer learning)                     & 70.66\% & - \\
Tiulpin,   et al., 2020  \cite{Tiulpin2020}         & OAI          & OAI               & MOST        & Random forest regression & Ensemble two models (Transfer learning)                 & -   & 66.69\%  \\
A.   Swiecicki et al., 2021 \cite{Swiecicki2021}          & MOST          & MOST              & MOST       & Faster   R-CNN                       & LAT   view                                                               & 61.18\% & - \\
A.   Swiecicki et al., 2021 \cite{Swiecicki2021}          & MOST          & MOST              & MOST       & Faster   R-CNN                       & PA   view                                                                & 70.85\%  & -\\
A.   Swiecicki et al., 2021 \cite{Swiecicki2021}          & MOST          & MOST              & MOST       & Faster   R-CNN                       & PA   and LAT channels                                                    & 71.90\% & - \\ \hline
Ours                             & MOST          & MOST              & MOST       & Yolo v3 tiny                         & Just   Classification                                                    & 71.93\%  & 72.51\%\\
Ours                             & MOST          & MOST              & MOST       & Yolo   v3 tiny                       & Segmentation based JSN                                                   & 58.86\% & - \\
Ours                             & MOST          & MOST              & MOST       & Yolo v3 tiny                         & Segmentation  based JSN+ Classification                                  & \textbf{75.86\%} & 73.11\% \\ 
Ours                             & MOST          & MOST              & OAI       & Yolo v3 tiny                         & Segmentation  based JSN+ Classification                                  & 64.48\% & \textbf{68.56\%} \\ \hline
\end{tabular}
\end{table*}

\begin{figure}[]
\centerline{\includegraphics[width = 0.8\columnwidth]{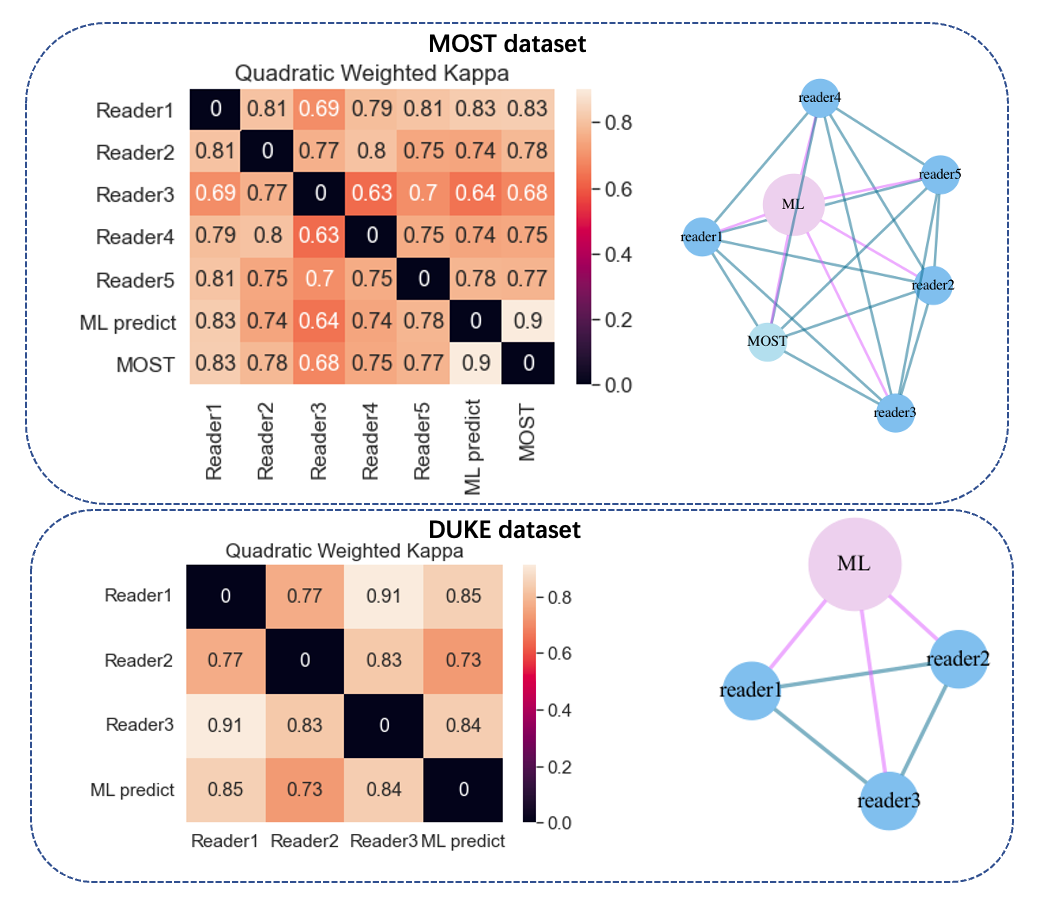}}
\caption{Agreement between the algorithm and radiologists measured by quadratic weighted Kappa (QWK) coefficient shown in the matrix as well as in a graphical form (distance between the nodes is inversely proportional to QWK), row 1: (left) is the QWK matrix for MOST dataset, (right) is the corresponding graphical form for the left matrix. Row 2: (left) is the QWK matrix for Duke dataset, (right) is the graphical form for Duke dataset.}
\label{fig11}
\end{figure}

The table of weighted quadratic-weighted Kappa ($\kappa$) is shown in \Cref{fig11} (left). 
On MOST dataset, the average quadratic weighted Kappa coefficient among all pairs of radiologists at our institution was 0.748 and it was 0.763 between our algorithm and the radiologists. 
The average $\kappa$ between Duke Hospital readers and MOST readers was 0.759, and the average $\kappa$ between our algorithm and MOST readers is 0.896. 
On Duke dataset, the average quadratic weighted Kappa among all the radiologists was 0.837, and it was 0.807 between our algorithm and the radiologists. 
By transferring the average $1 \\ {\kappa}$ into distances, we also drew a distance diagram as \Cref{fig11}(right). 
The diagram indicates that our algorithm showed an improved level of similarity to radiologists than radiologists between each other.

\section{Discussion}

We developed a novel multi-step deep learning model for automatic diagnosis of knee osteoarthritis through Kellgren-Lawrence grading. In contrast to previous studies, our model does not simply rely on a deep learning model to automatically make a knee OA classification from plain radiographs, but also incorporates bone segmentation, which is crucial to overall assessment accuracy. By utilizing an image segmentation model, we automatically calculated the JSN on both medial and lateral side and utilized them as supporting data for final decision. Compared to previous published studies, our model performs better in the predicted multi-class KL grading results and achieves a similar state-of-art level of OA instance ($KL\geq2$) detection.  Moreover, to our knowledge, this is the first study in OA detection to offer automatic knee segmentation as well as JSN grading visualization, which resulted in significantly improved accuracy and explainability since the segmentation masks and the JSN measurements can be displayed to the user.

Importantly, we made our software publicly available, both as code and as an executable, easy to use by those familiar with coding in python and those that have only rudimentary knowledge. The software, along with a user guide is available here: https://github.com/MaciejMazurowski/osteoarthritis-classification. We believe that the combination of the state-of-the-art performance of our algorithm, thorough evaluation with multiple datasets, and public availability of the code will allow for more rapid advances in OA imaging research. Our code could be applied to large repositories of OA cases for further research on the questions of surgery appropriateness and outcomes. To our knowledge there is no publicly available software of this type, until now.

As shown in \cref{table5}, our algorithm outperforms the state-of the-art methods in the ability to assess KL scores on the MOST dataset on both average accuracy and multi-class balanced accuracy, exceeding all the previously reported algorithms \cite{Tiulpin2020,Antony2017,Swiecicki2021}. Besides superior performance on MOST dataset, during the testing stage on OAI dataset, we also achieved a multi-class accuracy 64.48\% and a balanced accuracy 68.56\%. This setting shares a similarity as \cite{Tiulpin2020}, which is also trained on MOST dataset, and achieved a balanced accuracy at 66.71\% and 67.49\% on non-pretrained and pretrained models, respectively. However, though trained on MOST images, their model validation and selection are based on the OAI dataset performance, which might introduce a bias in our final reporting. Our model is developed only by supporting MOST training and validation sets with other subsets excluded, and the OAI test set indubitably indicates our method’s robustness to different artifacts and data acquisition settings. \cite{Thomas2020} reports a 70.66\% accuracy on OAI with a model that is directly trained on it. Our results are slightly lower, but this other method lacks our algorithm's detection explainability, and their method's lack of evaluation on additional datasets remains a potential concern for their model's generalization ability. Finally, compared with the method of \cite{Swiecicki2021}, our algorithm only requires PA view images, which is more applicable to radiologists' common practice.

To highlight the importance of detecting OA instances, our 2-class classification model combines (KL$\geq2$) for the presence of osteophytes and radiographic OA. As a result, our model achieves an accuracy of 0.93 and an F1-score of 0.94. This is higher than the reported score in other studies \cite{Antony2017,Tiulpin2020,1500491,biology10111107}. Moreover, \cite{1500491,biology10111107} methods all simplify the problem into binary classification only, i.e., they only focus on the determination of OA and abandon the discrimination of osteoarthritis severity, and we believe that our classification based on the 5 
grades of KL-score is more specific and clinically meaningful.

One consideration to be made is that the assessment of KL grades is subjective, and differents radiologists may have slight disagreements with one another, a phenomena known as inter-reader variability. Through analyzing the confusion matrix in Fig. 10, we observed that the highest difference between our algorithm and the MOST radiologist’s label is three degrees, and 93.52\% of the cases where our algorithm misguided suffers only 1-grade mismatch. 
The quadratic weighted Kappa of our method and MOST radiologists is 0.896. 
If we compare this to the average human agreement in KL grading (0.5-0.8) \cite{Tiulpin2018}, our method achieves a relatively high quadratic Kappa. 
Furthermore, from the comparison with the additional readers at our institution, we concluded that our algorithm performs at the level of a radiologist.
Since, the average kappa coefficient between our algorithm and the readers achieves a similar level as the average kappa coefficient among the Duke hospital readers, 0.763 for our algorithms and 0.748 among Duke readers on the MOST dataset, and 0.807 for our algorithm and 0.837 among Duke readers for Duke dataset. Our algorithm was placed at the center of the visualization mapping if we treat $1/\kappa$ as distance (\Cref{fig11} row 1 (right), showing a more substantial similarity of our algorithm with radiologists than the similarity between radiologists themselves. This indicates that our method can perform at a human level.

As described before, we observed an improvement in the segmentation-embedded algorithm’s performance compared with the single classification algorithm. This corresponds to our estimation that JSN plays a vital role in determining KL grades, both in machine learning and real-world diagnosis. An essential benefit of our method is the supplementary segmentation results and JSN grading, which could foster better reliability in clinical settings than other DL-based methods \cite{Tiulpin2018,Tiulpin2020,Antony2017,Thomas2020}. In addition, the software that we have made is publicly available with automatic processing of detection, segmentation, and classification, as well as a display of each step's results after loading raw DICOM images, making deployment immediately feasible. We believed that these features can provide further information and reduce the costs of routine work for further practitioners.

\section{Limitations}
There are a few limitations of our study. First, our model is only trained on the MOST dataset, and it may perform and generalize better if trained on a combination of other datasets, such as \cite{Tiulpin2018}, because the MOST images are collected by a standardized protocol, and therefore may not represent the entire distribution of possible applicable images. Including data from other institutions could result in an ever-higher generalizability.

Furthermore, in some of the misclassified images, our radiologists strongly disagreed with the ground truth KL grades in both OAI and MOST datasets. As such, it is possible that in certain cases in the training set, the ground truth KL grades were erroneous. It remains a question of how to get rid of these potentially misleading cases during training, or perform further study on these misclassified cases. Further studies could also consider incorporating additional OA features, such as medial tibial attrition, medial tibial sclerosis, and lateral femoral sclerosis.

\section{Conclusion}
In this paper, we a proposed five-step deep learning algorithm which uses segmentation and classification in parallel to achieve state-of-the art assessment performance of knee osteoarthritis severity at the radiologist level. Thorough evaluation with three datasets and a comparison to the performance of multiple experienced readers showed that our algorithm performs at the level of radiologists. Our software has been made publicly available and easy to use for further research.

\bibliographystyle{unsrtnat}
\bibliography{references}  






\end{document}